\documentstyle[preprint,aps]{revtex}
\begin{document}
\preprint{LA-UR-93-4317-REV, Accepted for publication in Phys. Lett. B}
\hsize = 7.0in
\widetext
\def\eq{\,=\,}
\def\beq{\begin{equation}}
\def\eeq{\end{equation}}
\def\l{\left(}
\def\r{\right)}
\def\po{\mathaccent 23 p}
\def\pp{p^+}
\def\pr{p_r}
\def\pl{p_\ell}
\draft
\title{Parity and Fermions in
 Front-Form: An Unexpected Result}

\author{Mikolaj Sawicki $^a$ and D. V. Ahluwalia $^b$
\footnotemark[1]\footnotetext[1]
{Internet: AV@LAMPF.LANL.GOV }}

\address{$^a$ Department of Physical Science,
John A. Logan College, Carterville,
Illinois 62918, USA}

\address {$^b$ Nuclear and Particle Physics Research Group, P-11,
  MS H-846, Los Alamos National Laboratory,\\
 Los Alamos, New Mexico 87545, USA }

\maketitle

\vskip 0.7in
\hrule\bigskip

\noindent
{\bf Abstract}
\bigskip

We show that under the operation of parity the {\it front-form} $(1/2,\,0)$ and
$(0,\,1/2)$ Weyl spinors (massive or massless) do not get interchanged. This
has the important consequence that if a front-form theory containing
$(1/2,\,0)\oplus(0,\,1/2)$ representation space has to be parity covariant then
one must study the evolution of a physical system not only along $x^+$ but also
along the $x^-$ direction. As a result of our analysis, we  find an indication
that there may be no halving of the degrees of freedom in the front form of
field theories.
\bigskip\hrule

\newpage
\noindent
{\bf 1. Introduction}
\bigskip

During a recent series of investigations
on the $(j,\,0)\oplus(0,\,j)$
representation space \cite{Ra,Rb,Rc,Rd,Re,Rf,Rg,Rh,Ri,Rj},
\footnotemark[1]
\footnotetext[1]
{\footnotesize
For instance, in Ref.
\cite{Ra} it was found that the $(1,0)\oplus(0,1)$ representation space
supports a Bargmann-Wightman-Wigner-type quantum field theory in which a boson
and its antiparticle have {\it opposite} relative intrinsic parity. In Ref.
\cite{Rd}, we applied the approach used previously for the instant-form
formalism \cite{Rc} to the front-form case and obtained the $(j,0)\oplus(0,j)$
spinors and generalized Melosh transformations for any spin. The front-form
formalism was seen to be endowed with several advantages.  The work of Ref.
\cite{Rd}, apart from the indicated generalization, reproduced some of the
well-known results of Melosh \cite{M}, Lepage and Brodsky \cite{LB}, and
Dziembowski \cite{D} for spin-$1/2$. Due to certain magic of Wigner's
time-reversal operator, in Ref. \cite{Ri} we were able to present a
Majorana-like construct in the $(j,0)\oplus(0,j)$ representation space.
} we have come across a problem in the front form of field theory with regard
to the operation of parity ($\cal P$). It is the purpose of this paper to bring
attention to this problem and present its solution. It is not the first time
that the problem of parity in the front form has been addressed and yet, as
will be seen, an {\it ab initio} analysis is needed. One of the earliest
considerations of this problem appears in the 1971 thesis of Soper
\cite{Soper}. More recently, Jacob \cite{OJ}  considered the question of
parity while considering the quantization of the scalar field, and he arrived
at similar conclusions to those presented in this paper in a different context.
We look at the transformation properties of the $(1/2,\,0)$ and $(0,\,1/2)$
Weyl spinors  in the front form and find that demand for $\cal P$-covariance
necessitates considering front-form evolution not only along $x^+$ (or $x^-$)
but simultaneously along $x^+$ and $x^-$. Our considerations are valid for
massive as well as massless particles, and are based upon space-time symmetries
that unambiguously establish the conclusions, thus disproving a widely-favored
conjecture regarding massive particles. Consensus, however, already exists on
McCartor's  results \cite{GM} for massless particles  where,  while
investigating the quantization of massless fields in the front form, he
concluded that a spin-$1/ 2$ system must be specified on  both $x^+$ and
$x^-$ surfaces. Our analysis can  be easily extended for all $(j,0)$ and
$(0,j)$ Weyl spinors and hence for all generalised Dirac-like
$(j,\,0)\oplus(0,\,j)$ fields. However, for conceptual clarity and general
familiarity, we choose spin-$1/2$ as an example case.

The present communication involves purely kinematical considerations. However,
kinematical considerations cannot be considered as devoid of dynamical
consequences. In general, if a theory is not $\cal P$-covariant kinematically
then dynamics cannot restore   $\cal P$-covariance. On the other hand, suppose
that,  we introduce an interaction via the principle of local gauge invariance
(without the $\openone\pm\gamma^5$ type projectors in the
$(1/2,\,0)\oplus(0,\,1/2)$ representation space) in a theory that is $\cal
P$-covariant at the kinematical level; then the resulting dynamical theory is
guaranteed to be $\cal P$-covariant as long as this covariance is not  violated
by the imposed boundary conditions \cite{cebaf}. It is in this context that the
kinematical considerations that follow are presented.

The remarks that we present are seemingly trivial, but in view of their
suggested relevance, we take the liberty of presenting them in this brief
essay.

\vskip 0.4in

\noindent
{\bf 2. Instant-Form Parity Transformation for Weyl and Dirac Spinors}
\bigskip

Within the above  framework, to define the problem, we recall  that in the {\it
instant form} $({1/2},\,0)$ and $(0,\,{1/ 2})$ Weyl spinors (for $m\not= 0$ as
well as $m=0$ ) transform as \cite{Ra,Rc,LR} \begin{eqnarray} &&\left({1/
2},\,0\right):\quad \phi_{_R}(p^\mu)\,=\, \exp\left[+\, {\bbox \varphi}\cdot
{{\bbox \sigma}\over 2} \right]\,\phi_{_R}({\overcirc p}^\mu)\quad,\nonumber\\
&&\left(0,\,{1/ 2}\right):\quad \phi_{_L} (p^\mu)\,=\,\exp\left[ -\, {\bbox
\varphi}\cdot{{\bbox \sigma}\over 2} \right]\,\phi_{_L}({\overcirc p}^\mu)
\quad. \label{if} \end{eqnarray} In Eqs. (\ref{if}), $p^\mu$ represents the
four-momentum of the particle and ${\overcirc p}^\mu$ corresponds to the
particle at rest. The boost parameter $\bbox \varphi$ that appears in Eq.
(\ref{if}) is defined
 as
\begin{equation} \cosh(\varphi\,)
\,=\,\gamma\,=\,{1\over\sqrt{1-v^2}}\,=\,{E\over m},\quad \sinh(
\varphi\,)\,=\,v\gamma\,=\,{| {\bf p}\, |\over m},\quad \hat{\bbox  \varphi}
\,=\,{ {\bf p} \over {|{ \bf p}\,|}},\quad
\varphi\,=\,\vert{\bbox\varphi}\vert\quad, \label{bp} \end{equation} with $\bf
p$ the three-momentum of the particle. It is immediately obvious from Eqs.
(\ref{if}) and (\ref{bp}) that under the operation parity, $\cal P$,
$({1/2},\,0)$ and $(0,\,{1/2})$ representation spaces  get interchanged,
\begin{equation} {\cal P}:\quad \left({1/2},\,0\right)\,\leftrightarrow\,
\left(0,\,{1/ 2}\right)\quad.\label{ifp} \end{equation} As is well known, it is
because of the result (\ref{ifp}) that any parity-covariant description
involving a spin-$1/2$ system must include both the $(1/2,\,0)$ and $(0,\,1/2)$
Weyl spinors. One of the easiest, and most familiar, way to incorporate the
result (\ref{ifp}) into a theory that includes  spin-$1/2$ fermions
is to introduce the $(1/2,\,0)\oplus(0,\,1/2)$ Dirac
spinor, which in the chiral representation (argument $p^\mu$ of spinors in the
chiral representation  are enclosed in curly brackets) reads
\begin{equation}
\psi\{p^\mu\}\,=\,
\left[
\begin{array}{c}
\phi_{_R}(p^\mu)\\
\phi_{_L}(p^\mu)
\end{array}\right]\quad.
\end{equation}
These results are well known and can indeed be found in any modern textbook
on
quantum field theory \cite{LR,MK,GS}. For the sake of later reference,
let's note that the  familiar \cite{BD} canonical representation
is defined as (argument $p^\mu$ of spinors in the canonical representation
are enclosed in square brackets)
\begin{equation}
\psi[\,p^\mu]\,=\,
{1\over{\sqrt{2}}}\,\left[
\begin{array}{ccc}
\openone &{\,\,\,}& \openone\\
\openone &{\,\,\,}& -\openone
\end{array}
\right]\,
\psi\{p^\mu\}\quad,\label{s}
\end{equation}
where $\openone$ is a $2\times 2$ identity matrix. Under the operation of
parity operator $S({\cal P})=\gamma^0$ in the $(1/2,\,0)\oplus(0,\,1/2)$
representation space, the particle and antiparticle spinors, in the usual
notation of Refs. \cite{Rc,BD}, transform as
\begin{eqnarray}
u_\sigma[\,p^{\prime\,\mu}] \,=\,+\,\gamma^0\,u_\sigma[\,p^\mu]\quad,\nonumber
\\
v_\sigma[\,p^{\prime\,\mu}]
\,=\,-\,\gamma^0\,v_\sigma[\,p^\mu]\quad.\label{ifuv}
\end{eqnarray}
The $p^{\prime\,\mu}$ is the parity-transformed $p^\mu$.

\vskip 0.4in
\noindent
{\bf 3. Front-Form Parity Transformation for Weyl and Dirac Spinors}
\bigskip

The front-form
counterpart of the simple and important instant-form relation (\ref{ifp}),
and other equations such as (\ref{ifuv}),  is a little subtler.
To see this, note that the counterpart of transformation properties of the
Weyl spinors in the front-form of evolution {\it associated with
$x^+=x^0+x^3$} reads (as was recently shown in Ref. \cite{Rd})
\begin{eqnarray}
&&\left({1/ 2},\,0\right)^{[x^+]}:\quad \phi^{[x^+]}_{_R}(p^\mu)\,=\,
\exp\left[+ \,{\bbox\beta}\cdot  {{\bbox \sigma}\over 2}
\right]\,\phi^{[x^+]}_{_R}({\overcirc p}^\mu)\quad,\nonumber\\
&&\left(0,\,{1/ 2}\right)^{[x^+]}:\quad \phi^{[x^+]}_{_L}
(p^\mu)\,=\,\exp\left[ - \,{\bbox\beta^\ast}
\cdot{{\bbox \sigma}\over 2}
\right]\,\phi^{[x^+]}_{_L}({\overcirc p}^\mu) \quad. \label{ff}
\end{eqnarray}
The superscript $[x^+]$ in the above equations serves the purpose of reminding
that these relations hold true for the evolution along $x^+$. The
${\bbox\sigma}$ are the standard Pauli matrices. The boost parameter
${\protect\bbox \beta}$ that appears in Eq. (\ref{ff})
is defined \cite{Rd}  as:
\begin{equation}
\protect\bbox{ \beta}
\,=\, \eta\,\left(\alpha\,v^r\,,\,\,-i\,\alpha\,  v^r\,,\,\,1\right)\quad,
\end{equation}
where
$
\alpha\,=\,\left[1\,-\,\exp(-\eta)\right]^{-1}\,,
$
 $v^r\,\,=\,\,v_x\,+\,i\,v_y$ (and
$v^\ell\,\,=\,\,v_x\,-\,i\,v_y$). In terms of the front-form variable
$p^+ \equiv E+p_z$, one can show that
\begin{equation}
\cosh(\eta/2)=\Omega
\left(p^+ + m\right)\,,\,
\sinh(\eta/2)=\Omega
\left(p^+ - m\right)\,,
\end{equation}
with
$
\Omega\,=\,\left[1/ (2 m)\right]\sqrt{m/ {p^+}}\,.
$
Under the operation of parity, $\cal P$,  an  inspection of Eqs. (\ref{ff}),
indicates that
$({1/2},\,0)$ and $(0,\,{1/2})$ representation
spaces do {\it not}  get interchanged;
\begin{equation}
{\cal P}:
\quad \left({1/2},\,0\right)^{[x^+]}\,\not\leftrightarrow\, \left(0,\,{1/
2}\right)^{[x^+]}\quad.\label{ffp}
\end{equation}
In fact we ourselves had failed to notice this in Ref. \cite{Rd}. (See the
paragraph after Eq. (15) of Ref. \cite{Rd}. However, the results obtained
in that article remain unaffected.)

To gain some physical understanding  of this result we note that under the
operation of parity, $\cal P$, unlike  the instant-form direction of evolution
$t\,=\,x^0$, the front-form direction of evolution that we picked above,
$x^+$, gets interchanged with $x^-$; and we should therefore obtain
counterparts of (\ref{ff}) and (\ref{ffp})  for the evolution along  the
parity-transformed $x^+$, that is $x^-$, and see how the Weyl spinors
transform. Algebraically this exercise is
not trivial, but it  parallels  our previous analysis of Ref.
\cite{Rd}. Here we just quote the result of our calculations. We find that the
$(1/2,\,0)$ and $(0,\,1/2)$ Weyl spinors in the front form of evolution {\it
associated with $x^-$} direction transform as
follows
\begin{eqnarray}
&&\left({1/ 2},\,0\right)^{[x^-]}:\quad \phi^{[x^-]}_{_R}(p^{\mu})\,=\,
\exp\left[+\, {\bbox\beta^\ast}\cdot  {{\bbox \sigma}\over 2}
\right]\,\phi^{[x^-]}_{_R}({\overcirc p}^\mu)\quad,\nonumber\\ &&\left(0,\,{1/
2}\right)^{[x^-]}:\quad \phi^{[x^-]}_{_L} (p^{\mu})\,=\,\exp\left[ -\,
{\bbox\beta} \cdot{{\bbox \sigma}\over 2} \right]\,\phi^{[x^-]}_{_L}({\overcirc
p}^\mu) \quad; \label{ffb}
\end{eqnarray}
and under the operation of parity, $\cal P$,
$({1/2},\,0)$ and $(0,\,{1/2})$ representation
spaces do {\it not}  get interchanged
\footnotemark[2]\footnotetext[2]
{\footnotesize The superscript $[x^-]$ in the above
equations serves the purpose
of reminding that these relations hold true for the evolution along $x^-$.}
\begin{equation}
{\cal P}:
\quad \left({1/2},\,0\right)^{[x^-]}\,\not\leftrightarrow\, \left(0,\,{1/
2}\right)^{[x^-]}\quad.\label{ffpextra}
\end{equation}

Comparison of transformation properties (\ref{ff}) and (\ref{ffb}) yields the
front-form counterpart of the instant-form relation (\ref{ifp}),
%\begin{equation}
%{\cal P}:\quad
%(1/2,\,0)^{[x^\pm]}\,\leftrightarrow \,(0,\,1/2)^{[x^\mp]}
%\end{equation}
\begin{equation}
{\cal P}:\quad
\left\{
\begin{array}{l}
(1/2,\,0)^{[x^+]}\,\leftrightarrow \,(0,\,1/2)^{[x^-]}\\
(0,\,1/2)^{[x^+]}\,\leftrightarrow \,(1/2,\,0)^{[x^-]}
\end{array}
\right.\quad.
\end{equation}
Thus, under the operation of parity, the representation space
$(1/2,\,0)^{[x^+]}\oplus (0,\,1/2)^{[x^+]}$ maps one-to-one onto
$(1/2,\,0)^{[x^-]}\oplus (0,\,1/2)^{[x^-]}$. To be more explicit, one may carry
out an exercise similar to the one presented in our recent work
\cite{Rd} and obtain the
$u^{[x^-]}_{\mu}[\,p^{\prime\,\mu}]$ and
$v^{[x^-]}_{\mu}[\,p^{\prime\,\mu}]$
spinors in the $(1/2,\,0)^{[x^-]}\oplus
(0,\,1/2)^{[x^-]}$ representation space.
To obtain $u^{[x^-]}_{\mu}[\,p^{\prime\,\mu}]$ and
$v^{[x^-]}_{\mu}[\,p^{\prime\,\mu}]$ the reader may find it convenient to
first rewrite (\ref{ffb}) as
\begin{eqnarray}
&&\left({1/ 2},\,0\right)^{[x^-]}:\quad \phi^{[x^-]}_{_R}(p^{\prime\mu})\,=\,
\exp\left[-\, {\bbox\beta^\ast}\cdot  {{\bbox \sigma}\over 2}
\right]\,\phi^{[x^-]}_{_R}({\overcirc p}^\mu)\quad,\nonumber\\ &&\left(0,\,{1/
2}\right)^{[x^-]}:\quad \phi^{[x^-]}_{_L} (p^{\prime\mu})\,=\,\exp\left[ +\,
{\bbox\beta} \cdot{{\bbox \sigma}\over 2} \right]\,\phi^{[x^-]}_{_L}({\overcirc
p}^\mu) \quad, \label{ffbextra}
\end{eqnarray}
so that the arguments on the left hand side are the parity transformed
$p^\mu$.
We already have  the explicit
expressions for the
$(1/2,\,0)^{[x^+]}\oplus (0,\,1/2)^{[x^+]}$ spinors,
$u^{[x^+]}_h[\,p^\mu]$ and
$v^{[x^+]}_h[\,p^\mu]$, from Ref. \cite{Rd}. The parity operator,
$S({\cal P})\,=\,\gamma^0$, remains unaltered in going from
the instant form to the front form
because Melosh transformation, as was
explicitly proved in \cite{Rd}, does not mix particle and antiparticle spinors,
{\it and} the front-form $(1/2,\,0)\oplus(0,\,1/2)$ spinors turn out to be the
superposition of the instant-form spinors with $p^\mu$-dependent coefficients
contained in the Melosh matrix for spin-$1/2$. The above indicated exercise
yields the front-form counterpart of the identities (\ref{ifuv}),
\begin{eqnarray}
u^{[x^-]}_{\mu}[\,p^{\prime\,\mu}]
\,=\,+\,\gamma^0\,u^{[x^+]}_h[\,p^\mu]\quad,\nonumber \\
v^{[x^-]}_{\mu} [\,p^{\prime\,\mu}]
\,=\,-\,\gamma^0\,v^{[x^+]}_h[\,p^\mu]\quad.\label{ffuv}
\end{eqnarray}
The $h $ and $\mu$ in the above expressions correspond to the helicity
degrees of freedom associated with the front-form helicity operators
(respectively associated with evolution along $x^+$ and $x^-$):
\begin{eqnarray}
{\cal J}^{[x^+]}_3 \equiv J_3 \,+\,{1\over
P_-}\left(G_1\,P_2\,-\,G_2\,P_1\right)\quad,
\\
{\cal J}^{[x^-]}_3 \equiv J_3 \,+\,{1\over P_+}\left(D_1\,P_2\,-\,D_2\,
P_1\right)\quad.
\end{eqnarray}
For dynamical significance and  the definition of various generators involved
in the above expressions, we refer
 the reader  to Sec. II of Ref. \cite{Rd}.
The  non-trivial space-time structure of the results we obtain, such as
Eqs. (\ref{ffuv}), is  intuitively satisfactory.

\vskip 0.4in
\noindent
{\bf 4. Concluding Remarks}
\bigskip

The main result of this paper is to show that under the operation of parity the
{\it front-form} $(1/2,\,0)$ and $(0,\,1/2)$ Weyl spinors (massive or massless)
do not get interchanged. As a consequence, if we only consider  $x^+$, or
equivalently $x^-$, as the ``front-form time'' for the front-form evolution
then such a description cannot be parity covariant; and any gauge interactions
introduced for such a system would {\it necessarily} result in a
parity-violating dynamical system ({\it cf.} comments made towards the end of
the Introduction) . Therefore, if a front-form theory containing
$(1/2,\,0)\oplus(0,\,1/2)$ representation space has to be parity covariant then
one must study the evolution of a physical system not only along $x^+$ but also
along the $x^-$ direction.

Precisely how such a evolution is to be implemented is not yet fully clear.
However, we venture a few preliminary remarks. Following the pioneering work
of Chang, Root and Yan  \cite{CRY} we know  that  only half of the degrees of
freedom associated with the field $\Psi^{[x^+]}(x)$ constructed from
$u^{[x^+]}_h[\,p^\mu]$ and $v^{[x^+]}_h[\,p^\mu]$ are dynamical. Similarly,
only half of the degrees of freedom associated with the field
$\Psi^{[x^-]}(x)$, constructed from $u^{[x^-]}_\mu[\,p^\mu]$ and
$v^{[x^-]}_\mu[\,p^\mu]$, are dynamical. Under the operation of parity, using
(\ref{ffuv}), we get:  $\Psi^{[x^+]}(x)\, \leftrightarrow \,\Psi^{[x^-]}(x)$.
Each of the  $\Psi(x)$ field carries two independent dynamical degrees of
freedom (related via operation of parity). Since parity covariance demands that
we include $\Psi^{[x^+]}$ as well as $\Psi^{[x^+]}$ in the theory we have four
(same as in instant form)
dynamical degrees of freedom.

In the context of these tentative remarks on loss of degrees of freedom we note
that many physicists feel uncomfortable with the halving of degrees of freedom
in the front-form of field theories. Already in 1963, a decade before
front-form field theory became popular with nuclear physicists, Penrose [see
Ref. \cite{RP}, p. 235, footnote \# 15] in a similar context had noted ``This
explanation leaves something to be desired, however. ... There is probably a
subtler reason for this halving of the initial data functions.'' As a result of
our analysis, we now find an indication that there may be no halving of the
degrees of freedom in the front form of field theories.
\footnotemark[3]\footnotetext[3] {After this work was completed, Jacob sent us
a preprint \cite{OJb} where he shows that the result (\ref{ffuv}),
obtained by us, is essential for front-form quantization, which involves
specification of a system on both $x^+=0$ and $x^-=0$ surfaces.}

Finally, a very different perspective on quantum field theories in the front
form and our unexpected results can be gained by realising that light-like
surface is a {\it characteristic} surface of  the Klein-Gordon and Dirac
equations. As a result, as Ligterink and Bakker \cite{Ligt} note ``The
mathematical theory of partial differential equations tells us some of the
strange effects which might be expected if we use a light-like surface as
boundary on which we define the initial field ...  . The solutions  of
differential equations have a number of special properties, often independent
of the fact that we have {\it massive} or {\it massless } fields... .''

\vskip 0.4in \noindent {\bf 5. Acknowledgements} \bigskip

It is our pleasure to thank Ovid  Jacob  (SLAC)  for insightful conversations
on the general subject of this work and George Kahrimanis  for helpful
comments.

The work of D. V. Ahluwalia  was done under the auspices of the U. S.
Department of Energy.


\begin{references}

\bibitem{Ra} D. V. Ahluwalia, M. B. Johnson, and T. Goldman, Phys. Lett. B  316
(1993) 102.

\bibitem{Rb} D. V. Ahluwalia and T. Goldman, Mod. Phys. Lett. A
 8 (1993) 2623.

\bibitem{Rc} D. V. Ahluwalia and D. J. Ernst,
Int. J. Mod. Phys. E  2  (1993) 397.

\bibitem{Rd} D. V. Ahluwalia and M. Sawicki,
Phys. Rev. D  47 (1993) 5161.

\bibitem{Re} D. V. Ahluwalia and D. J. Ernst,
Phys. Rev. C  45 (1992) 3010.

\bibitem{Rf} D. V. Ahluwalia and D. J. Ernst,
Mod. Phys. Lett. A  7 (1992) 1967.

\bibitem{Rg} D. V. Ahluwalia and D. J. Ernst,
Phys. Lett. B 287 (1992) 18.

\bibitem{Rh} D. V. Ahluwalia,
Phys. Lett. B  277 (1992)  243.

\bibitem{Ri} D. V. Ahluwalia, T. Goldman, and M. B. Johnson,
Mod. Phys. lett. A 9 (1994) 439.

\bibitem{Rj} D. V. Ahluwalia, M. B. Johnson, and T. Goldman,
(to appear) {\it in} ``Proceedings of the III International Wigner Symposium,''
Christ Church, Oxford, September 1993.


\bibitem{M} H. J. Melosh,
Phys. Rev. D 9 (1974) 1095.

\bibitem{LB} G. P. Lepage and S. J. Brodsky,
Phys. Rev. D 2157 (1980) 22.

\bibitem{D} Z. Dziembowski,
Phys. Rev. D  37 (1987) 768.

\bibitem{Soper} D. E. Soper,
SLAC Report 137, 1971 (unpublished).

\bibitem{OJ} O. C.  Jacob,
SLAC Preprint 6391, 1993.

\bibitem{GM} G. McCartor, Z. Phys. C  41 (1988) 271; and private communication
(1993); G. McCartor and D. G. Robertson, Z. Phys. C  53 (1992) 679.

\bibitem{cebaf} D. V. Ahluwalia, in {\it Proceedings of the HUGS at CEBAF},
Newport News, 1990, edited by W. W. Buck (Continuous Electron Beam Accelerator
Facility and Hampton University, Newport News, 1990).


\bibitem{LR} L. H. Ryder,
{\it Quantum Field Theory} (Cambridge University Press, Cambridge, England,
1987).

\bibitem{MK} M. Kaku,
{\it Quantum Field Theory} (Oxford University Press, Oxford, England, 1993).

\bibitem{GS} G. Sterman,
{\it An Introduction to Quantum Field Theory}
(Cambridge University Press, Cambridge, England, 1993).


\bibitem{BD} J. D. Bjorken and S. D. Drell,
{\it Relativistic Quantum Mechanics} (McGraw-Hill Book Co., New York, 1964).



\bibitem{CRY} S. J. Chang, R. G. Root, and T. M. Yan,
Phys. Rev. D  7 (1973) 1133.

\bibitem{RP} R. Penrose, Aerospace Research Laboratories pre-print 63-56,
ed. P. G. Bergmann (1963); reprinted as a ``Golden Oldie'' in Gen. Rel.
Grav.  12 (1980) 225.

\bibitem{OJb} O. C. Jacob,
SLAC Preprint 6392, 1993; and private communications (1993,1994).


\bibitem{Ligt} N. E. Ligterink and B. L. G. Bakker, Unnumbered preprint
entitled (1993) ``Partial Differential Equations and Null-Plane Field Theory.''
Also see, L. Chao, Mod. Phys. Lett. 8 (1993) 3165; R. K. Sachs,
J. Math. Phys. 3 (1962) 908; H. M. Z. Hagen and H.-J. Seifert,
Gen. Rel. Grav. 8 (1977) 259.


\end{references}
\end{document}